\documentclass{article}
\usepackage{spconf,amsmath,graphicx}
\usepackage{enumitem}
\setlist{nosep, leftmargin=14pt}
\usepackage{amsfonts}
\usepackage{mwe}

\title{Benchmarking 3D multi-coil NC-PDNet MRI Reconstruction}
\name{A. Tanabene $^{1,}$$^{2,}$$^3$, Chaithya G. R. $^{2,}$$^3$, A. Massire $^{1}$, M. Nadar $^{4}$, P. Ciuciu $^{2,}$$^3$.}
\address{$^1$Siemens Healthineers, Courbevoie, 92400, France\\
	$^2$CEA, NeuroSpin, CNRS, Université Paris-Saclay, Gif-sur-Yvette, 91191, France\\
	$^3$Inria, MIND, Palaiseau, 91120, France\\
        $^4$Siemens Healthineers, Princeton, 08540, NJ, USA}

\begin{document}
\ninept
\maketitle
\begin{abstract}
Deep learning has shown great promise for MRI reconstruction from undersampled data, yet there is a lack of research on validating its performance in 3D parallel imaging acquisitions with non-Cartesian undersampling. In addition, the artifacts and the resulting image quality depend on the under-sampling pattern. To address this uncharted territory, we extend the Non-Cartesian Primal-Dual Network (NC-PDNet), a state-of-the-art unrolled neural network, to a 3D multi-coil setting. We evaluated the impact of channel-specific versus channel-agnostic training configurations and examined the effect of coil compression. Finally, we benchmark four distinct non-Cartesian undersampling patterns, with an acceleration factor of six, using the publicly available Calgary-Campinas dataset. Our results show that NC-PDNet trained on compressed data with varying input channel numbers achieves an average PSNR of 42.98dB for 1 mm isotropic 32 channel whole-brain 3D reconstruction. With an inference time of 4.95sec and a GPU memory usage of 5.49 GB, our approach demonstrates significant potential for clinical research application.

\end{abstract}
\section{Introduction}
\label{sec:intro}
Magnetic Resonance Imaging (MRI) is a non-invasive medical imaging technique that offers excellent soft tissue contrast. Advancements in MRI technology are constantly evolving to achieve higher image resolution and acquisition speed, enabling faster, more detailed, and more accurate imaging. In the quest for high SNR and high isotropic spatial resolution images, 3D parallel-imaging protocols are increasingly used, often resulting in longer scan times. Inspired by compressed sensing theories~\cite{Lustig2007}, accelerated scans can be performed by subsampling the k-space in a variable density. The fully 3D non-Cartesian imaging paradigm allows us to attain such variable density sampling, with denser sampling at the center of the k-space and tapering off towards the periphery, offering faster and more efficient coverage of the k-space than Cartesian sampling~\cite{wright2014non}. However, while acquisition speed increases, this change also transfers computation complexity and time requirements to the image reconstruction task.

Recently, deep learning has emerged as an alternative for fast MR image reconstruction without compromising image quality. Among the most promising solutions in this field are physics-driven neural networks that unroll optimization algorithms \cite{hammernik2018learning,zeng2021review,adler2018learned,heckel2024deep} and produce enhanced MR images from raw k-space data. However, most recent deep neural networks do not effectively scale to 3D multi-coil acquisition setups. For example, the non-Cartesian primal-dual network (NC-PDNet)~\cite{ncpdnet}, the first density-compensated unrolled neural network designed for non-Cartesian imaging, has only been validated on multi-coil 2D and single-coil magnitude-only 3D k-space data.

Training end-to-end unrolled neural networks on 3D multi-coil non-Cartesian data presents significant challenges. One major issue is that handling non-Cartesian data requires using the Non-Uniform Fast Fourier Transform (NUFFT) operator, which is more computationally and memory-intensive than the standard Fast Fourier Transform, especially as the number of channels increases. Furthermore, incorporating 3D convolutional neural networks (CNN) into unrolled architectures requires managing a much larger number of parameters~(i.e. neurons) than in 2D configurations. These high memory requirements have significantly constrained research efforts in scaling recent advances to 3D multi-coil non-Cartesian settings. To manage this, memory-efficient solutions, such as coil compression, are recommended to reduce the input size of 3D multi-coil k-space data.

In this work, we address the challenge of extending NC-PDNet to non-Cartesian 3D multi-coil MRI reconstruction. Further, to understand the artifacts and reconstructed image quality, we conduct a retrospective ablation study with four different non-Cartesian undersampling patterns. Finally, we evaluate the impact of coil compression and compare channel-specific versus channel-agnostic training approaches.

\section{Experimental Setup}
\subsection{Model}
\label{sec:format}
In our multi-coil non-Cartesian under-sampling context, the reconstructed image is recovered by solving the following ill-posed optimization problem:
\begin{equation}
\label{eq:inverse_prob}
\underset{\boldsymbol{x} \in \mathbb{C}^N}{\arg \min } \frac{1}{2L}\sum_{\ell=1}^L \left\|\boldsymbol{y}_{\ell}-\mathcal{F}_{\Omega} \boldsymbol{S}_{\ell} \boldsymbol{x}\right\|_2^2+\mathcal{R}(\boldsymbol{x})
\end{equation}
where $\mathcal{F}$ is the multi-coil 3D NUFFT with locations defined by sampling pattern ${\Omega}$ applied on $L$ number of coils. $\boldsymbol{S}_{\ell} $ refers to the sensitivity map of the ${\ell^{th}}$ corresponding coil, and $\mathcal{R}(\boldsymbol{x})$ serves as a prior for the solution, which is conventionally sparsity enforced in wavelet domain. $\boldsymbol{y}_{\ell} \in \mathbb{C}^M $ are the k-space measurements and $\boldsymbol{x} \in \mathbb{C}^N $ is the target image, where $M << N$.
NC-PDNet unrolls the proximal gradient descent, an iterative algorithm that solves our optimization problem~\eqref{eq:inverse_prob}, as follows:

\begin{align}
    x_{n+1} &= x_n - \epsilon_n \mathcal{A}^H \left( \mathcal{A} x_n - \boldsymbol{y} \right), \nonumber \\
    x_{n+1} &= \operatorname{prox}_{\epsilon_n \mathcal{R}} \left( x_{n+1} \right)
\end{align}
where with $\otimes$ being the tensor product,
\begin{equation*}
    \mathcal{A} = \left( \boldsymbol{I}_L \otimes \mathcal{F}_{\Omega} \right) \mathbb{S}, \quad
    \mathbb{S} = \begin{bmatrix} \boldsymbol{S}_1 \\ \vdots \\ \boldsymbol{S}_L \end{bmatrix}, \quad
    \boldsymbol{y} = \begin{bmatrix} \boldsymbol{y}_1 \\ \vdots \\ \boldsymbol{y}_L \end{bmatrix}.
\end{equation*}

\begin{figure}[htb]
    \centering
    \includegraphics[width=7.5cm]{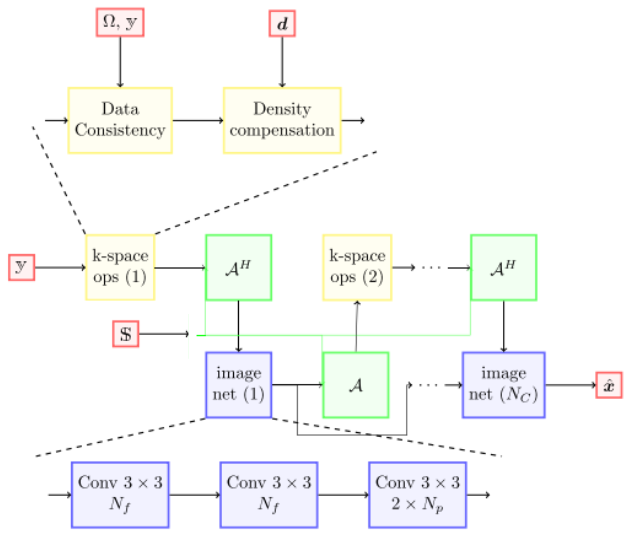}
    \vspace*{-.3cm}
    \caption{NC-PDNet architecture \cite{ncpdnet} for multi-coil MRI reconstruction, where \( N_f \) is the number of convolutional filters, \( N_P \) is the buffer size, and \( N_C \) is the number of unrolled iterations.}
    \label{fig:ncpdnet}
\end{figure}

The NC-PDNet architecture is made up of multiple blocks or iterations as shown in \ref{fig:ncpdnet}. We carry a buffer \( \boldsymbol{x}_b \) through iterations rather than relying on a single estimate. In each block $\boldsymbol{i}$, we first compute a data consistency~(DC) term in the k-space as $\boldsymbol{y}_{DC} = \mathcal{A}\boldsymbol{x}_{b}[0] - \boldsymbol{y}$, followed by density compensation $\boldsymbol{x}_{DC} = \mathcal{A}^H \boldsymbol{d} \boldsymbol{y}_{DC}$. A refinement network is then applied in the image domain using a residual CNN with three consecutive convolutions and ReLU activation, represented by $\boldsymbol{x}_{i+1} = \boldsymbol{x}_{i} + f_{\theta_i}(\boldsymbol{x}_{b}, \boldsymbol{x}_{DC})$. 

The model is trained end-to-end, with trainable layers of each block $f_{\theta_i}$ having distinct weights that are not shared across iterations. For this study, we used an optimized configuration for NC-PDNet to fit the model on a single GPU, with six unrolled iterations, a buffer size of 2, and 16 convolution filters. 
\subsection{Data}
In this work, we used raw data from the Calgary-Campinas dataset \cite{beauferris2022multi}, which includes 167 3D T1-weighted gradient-recalled echo scans with 1 mm isotropic sagittal acquisitions collected on a clinical 3-T MRI scanner from healthy subjects. The data set contains 117 scans (70.0\%) acquired with a 12-channel receiver coil and 50 scans (30.0\%) with a 32-channel coil. The acquisition parameters were TR / TE / TI = 6.3ms / 2.6ms / 650ms (93 scans, 55.7\%) or TR/TE/TI = 7.4ms / 3.1ms / 400ms (74 scans, 44.3\%). The matrix size was $N_x \times N_y \times N_z$ = 256 x 218 x [170–180], with the slice-encoded direction (kz) partially sampled up to 85\% of its matrix size. 

Subsequently, retrospective under-sampled data were obtained by projecting multi-coil reference images onto non-Cartesian under-sampled trajectories using the NUFFT operator. The ground truth was the square root of the sum of squares of all multi-coil Cartesian-sampled images. We randomly split the data into training~(70\%), validation~(10\%), and test (20\%)  sets. To perform coil compression, we applied singular value decomposition to the raw k-space data, retaining the singular vectors that account for 99\% of the cumulative explained variance. Accordingly, we reduced the 12-channel and 32-channel k-space data to four and seven compressed channels respectively.
 
\subsection{Non-Cartesian Sampling Patterns}

\begin{figure}[htb]

\begin{minipage}[b]{.48\linewidth}
  \centering
  \centerline{\includegraphics[width=3cm]{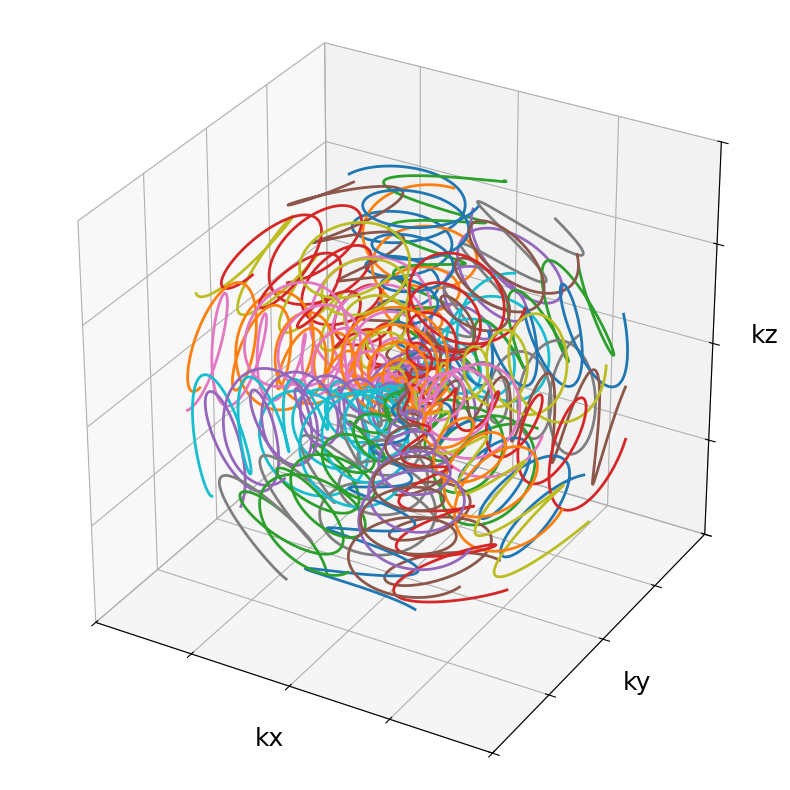}}
  \centerline{(a) 3D Cones}
\end{minipage}
\hfill
\begin{minipage}[b]{0.48\linewidth}
  \centering
  \centerline{\includegraphics[width=3cm]{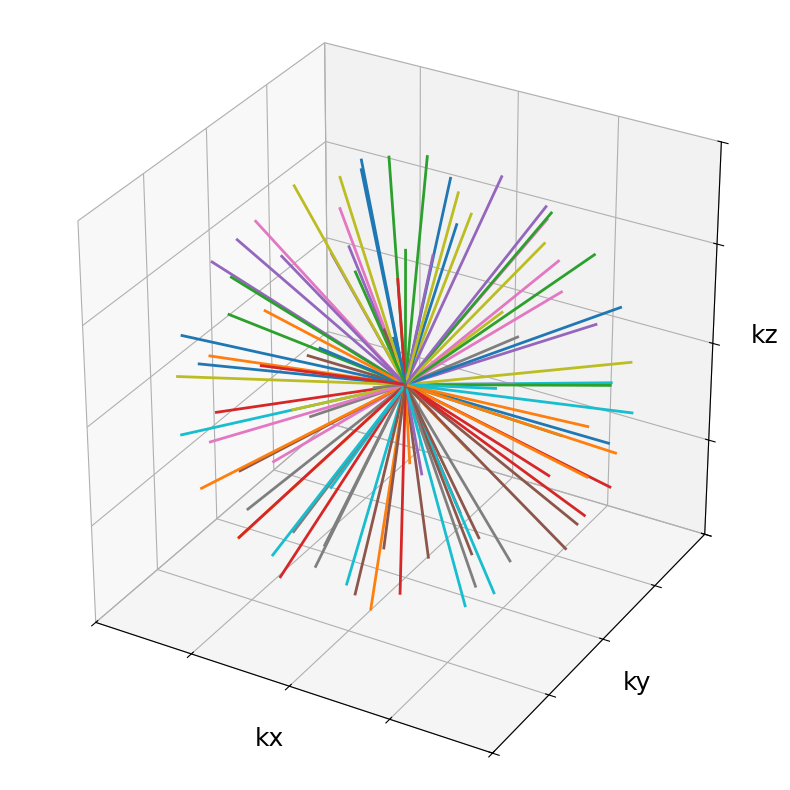}}
  \centerline{(b) 3D Radial}
\end{minipage}
\begin{minipage}[b]{.48\linewidth}
  \centering
  \centerline{\includegraphics[width=3cm]{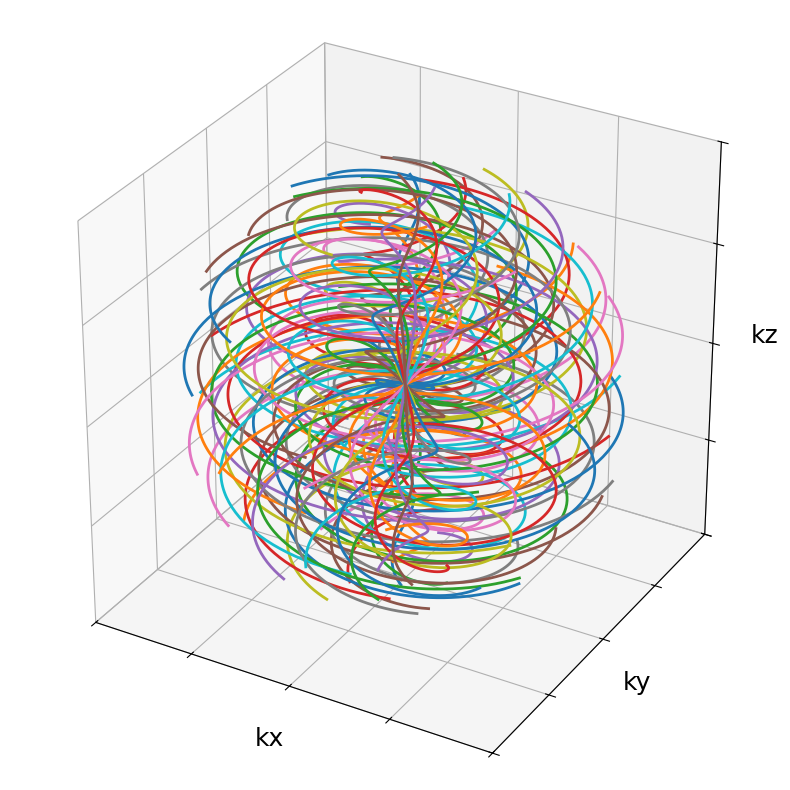}}
  \centerline{(c) TPI}
\end{minipage}
\hfill
\begin{minipage}[b]{0.48\linewidth}
  \centering
  \centerline{\includegraphics[width=3cm]{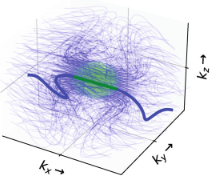}}
  \centerline{(d) GoLF-SPARKLING}
\end{minipage}
    \vspace*{-.3cm}
\caption{Trajectories are shown with a reduced number of shots for clarity. GoLF-SPARKLING combines a non-Cartesian SPARKLING portion (blue) with a grid-sampled low-frequency region (green).}
\label{fig:trajectories}
\end{figure}

Different 3D trajectories were generated to match an acceleration factor  of 6 ($\approx \frac{218\times170}{6}$ trajectories with 512 samples each), which results in scan time of nearly 1.5min for 1 mm isotropic whole brain MRI. For our comparison, a 3D radial trajectory with golden means-based structure \cite{3d_radial_golden_means} and a 3D cone trajectory \cite{cones}, both with center-out shots, were used. Also, as the density of the radial trajectories drops drastically at higher frequencies, we compared the reconstructed image quality using twisted projection imaging (TPI)~\cite{boada1997fast}. Finally, we also tested the reconstructions on the GoLF-SPARKLING \cite{giliyar2023improving} sampling trajectory, which extends the original SPARKLING approach~\cite{chaithya_optimizing_2022} that was originally tested for T2*-weighted imaging.
 In GoLF-SPARKLING, each sampling trajectory passes through the center of k-space as a Cartesian line, resulting in Cartesian sampling at the center of k-space, shown in green in Fig.~\ref{fig:trajectories}. 
 In practice, the acceleration factor of the GoLF-SPARKING trajectory was slightly higher (6.7) and had almost 42\% of the center of the k space sampled in a Cartesian way.
 
\subsection{Implementation, training and evaluation details}
All the code was implemented in PyTorch. Alternating between the k-space and image domains was achieved using forward and adjoint NUFFT operators. To address memory constraints, we used the MRI-NUFFT package \cite{comby_mri-nufft_2024} as its NUFFT implementations are memory and computationally efficient and support PyTorch tensors for the \texttt{CufiNufft} and \texttt{GpuNUFFT} backends. 

We used the mean absolute error (MAE) as the loss function, the Adam optimizer with a learning rate of $1e-3$, and a reduce-on-plateau scheduler. The training was carried out for 200 epochs for all the networks on a single A100 GPU with 80GB of VRAM. We evaluated the model's performance on the test set using the checkpoint giving the best validation scores. 

To assess the impact of channel configuration, we first conducted two separate training sessions for each non-Cartesian undersampling pattern: one with only 12-channel data and the other with 32-channel data. We refer to this approach as \emph{channel-specific} training. In addition, we conducted training on the entire dataset, which is designed as \emph{channel-agnostic} training.

The metrics used for evaluating and comparing the networks are the peak signal-to-noise ratio (PSNR) and structural similarity index (SSIM).

\section{Results and Analysis}
\subsection{Channel-Specific vs. Channel-Agnostic Training}
\begin{figure}[h!]
  \centering
  \centerline{\includegraphics[width=9cm]{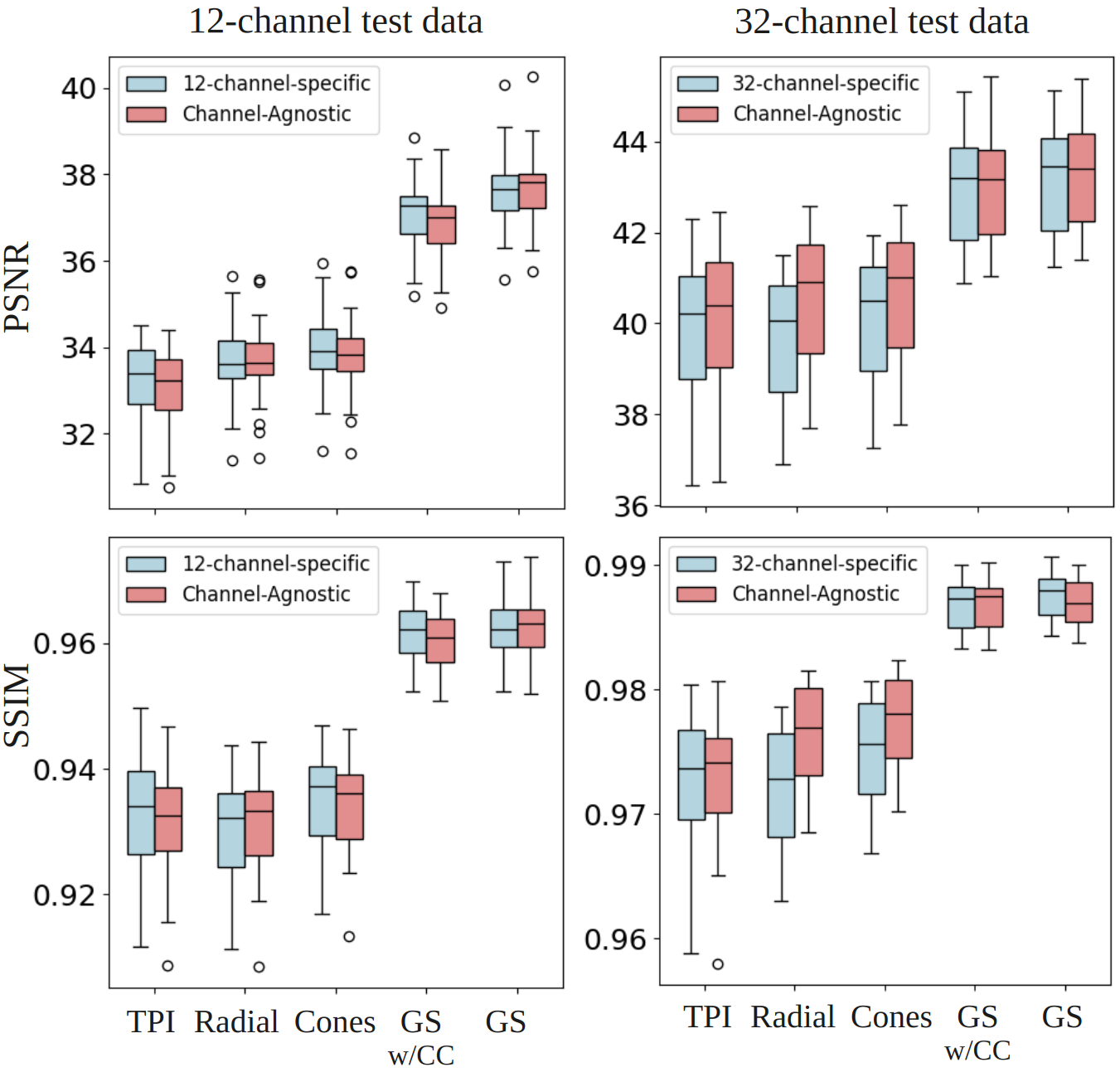}}
  \vspace*{-.5cm}
    \caption{Quantitative results of NC-PDNet with different trajectories, ordered from left to right: TPI, 3D Radial, 3D Cones, GoLF-SPARKLING (GS w/CC), and GoLF-SPARKLING without coil compression (GS).}
    \label{fig:boxplots}
\end{figure}

Box plots in Fig.~\ref{fig:boxplots} illustrate that channel-agnostic models, when evaluated on the 32-channel test subset, outperform the 32-channel-specific model trained and tested exclusively on 32-channel data, showing higher PSNR and SSIM scores. Meanwhile, the channel-agnostic model achieves comparable or slightly lower scores when tested against the 12-channel-specific model on 12-channel data. These findings suggest that training a single channel-agnostic model across all data is more practical and efficient, as it provides near-equivalent performance across both 32- and 12-channel configurations without the need for separate, channel-specific models.

\begin{table}[ht]
    \caption{Mean PSNR/SSIM scores of NC-PDNet with channel-agnostic training on 12-channel and 32-channel test subsets.   
         \vspace{0.3cm}
    \label{tab:example}}
    \begin{tabular}{|c|c|c|} 
        \hline
    \centering
        \textbf{Trajectory} & \textbf{12-ch. PSNR/SSIM} & \textbf{32-ch. PSNR/SSIM}\\ \hline
        TPI & 33.03 / 0.932  & 40.03 / 0.972\\ \hline
        3D Radial & 33.65 / 0.931 & 40.49 / 0.976 \\ \hline
        3D Cones & 33.82 / 0.934 & 40.55 / 0.977\\ \hline
        GS w/CC & \underline{36.83 / 0.960} & \underline{42.89 / 0.987} \\ \hline
        GS    & \textbf{37.62 / 0.962} & \textbf{43.33 / 0.987} \\ \hline
    \end{tabular} 
\end{table}

\subsection{Performance benchmarking of selected non-Cartesian sampling patterns}

\textbf{Quantitative results.} Results in Tab.~\ref{tab:example} show that the undersampling with the GOLF-SPARKLING~(denoted GS in the table, with or without coil compression~-- w/CC) outperforms all other methods, while the 3D cone trajectory performs slightly better than the 3D radial and TPI patterns. 
GoLF-SPARKLING achieves higher PSNR and SSIM scores due to its design, which incorporates Cartesian sampling in the central region of k-space. This sampling pattern enables a more accurate estimation of sensitivity maps, extracted by applying the adjoint NUFFT to a density-compensated low-frequency central region of k-space. As a result, GoLF-SPARKLING allows us to achieve superior performance compared to fully non-Cartesian sampling patterns.

\begin{figure*}[htb]
    \centering
    \includegraphics[width=1.0\textwidth]{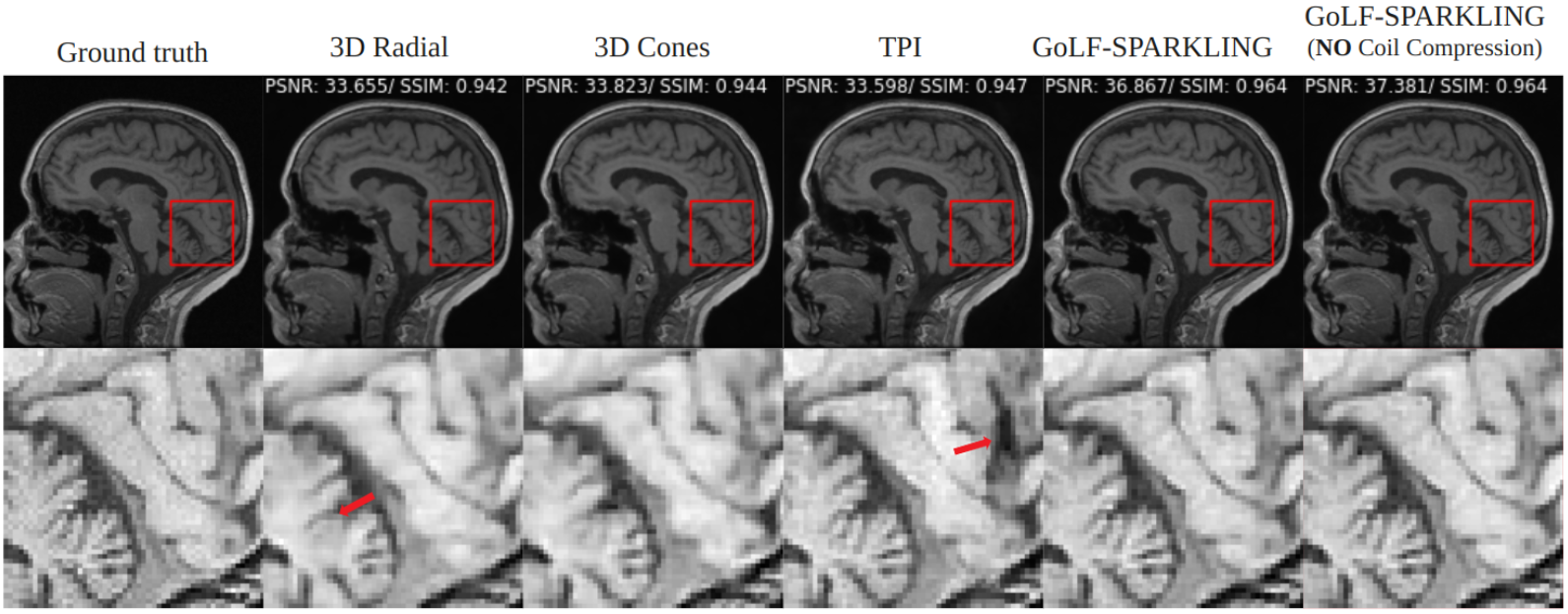}
    \vspace*{-.7cm}
    \caption{ Reconstruction results of the $90^{th}$ slice of file \texttt{e14079s3\_P09216.7} from the test set. The top row shows reconstructions by different methods, while the bottom row displays zoomed-in regions outlined by red frames. Volume-wise PSNR and SSIM scores are indicated in the top left corner of each image.\label{fig:fig3}}
\end{figure*}

\textbf{Qualitative results.} Visual evaluation of reconstructed MR images confirms the quantitative metrics. As illustrated in Fig.~\ref{fig:fig3}, the fine structures of the cerebellum are sharper and more accurately reconstructed using the GoLF-SPARKLING and TPI trajectories, in contrast to the blurriness observed with the radial and cone patterns. Additionally, GoLF-SPARKLING demonstrates a higher fidelity in recovering brain structures than TPI, which displays a darkened appearance in some areas relative to the ground truth. This underestimation of signal intensity in TPI could obscure finer details in specific regions, potentially affecting diagnostic interpretation.

\begin{figure}[h!]
\begin{minipage}[b]{1.0\linewidth}
  \centering
  \centerline{\includegraphics[width=8.5cm]{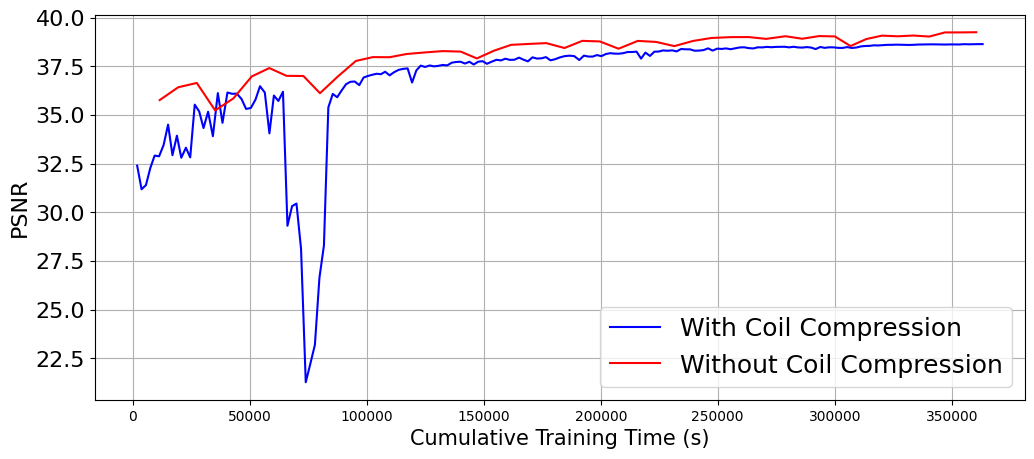}}
  \centerline{(a) Mean Test PSNR Score vs. Cumulative Training Time}\medskip
\end{minipage}
\begin{minipage}[b]{1.0\linewidth}
  \centering
  \centerline{\includegraphics[width=8.5cm]{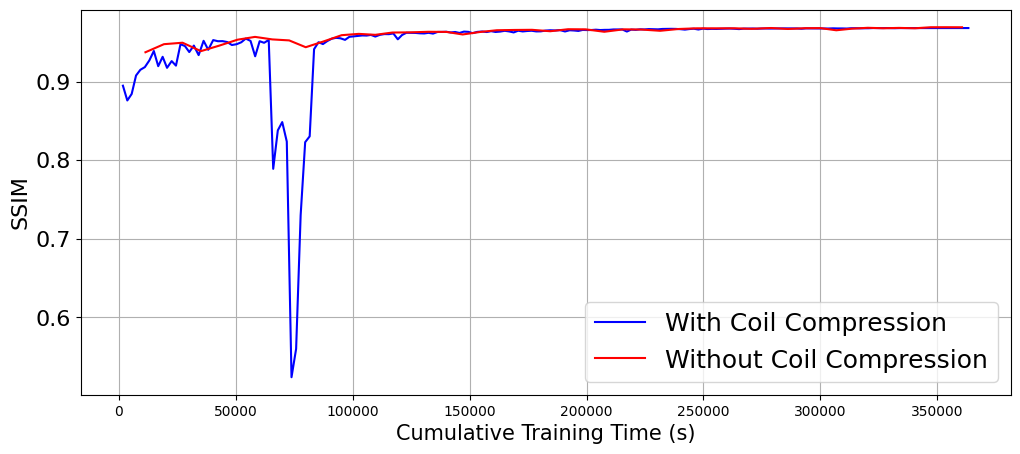}}
  \centerline{(b) Mean Test SSIM Score vs. Cumulative Training Time}\
\end{minipage}
\vspace*{-.9cm}
\caption{Evolution of test scores over cumulative training time for channel-agnostic NC-PDNet training.}
\label{fig:res}

\end{figure}
\subsection{Assessing the Impact of Coil Compression}
Plots of SSIM and PSNR test scores over the cumulative training time (Fig.~\ref{fig:res}) show that, for the same training duration, NC-PDNet with GS channel-agnostic training without coil compression achieves an equivalent average SSIM score and an increase of approximately +0.45dB in average PSNR compared to the same training performed with coil compressed k-space input data. In addition, coil compression training is less stable, as illustrated by the significant drop in performance in Fig.~\ref{fig:res}. 
However, Fig.~\ref{fig:fig3} shows a very close perceived visual quality between the reconstructed images of models trained with and without coil compression, suggesting that in case of memory constraints, coil compression is a key strategy to maintain good performances while allowing for scalability with respect to the dimension of the multi-coil input data set.

\begin{table}[ht]
    \centering
    \caption{Comparison of NC-PDNet with GS, trained with and without coil compression~(w/CC and w/o CC, resp.), showing average epoch training time and peak GPU memory usage per training step. Training data: 116 volumes (69.8\% 12-channel, 30.2\% 32-channel).    \label{tab:training}}
         \vspace{0.cm}
    \begin{tabular}{|c|p{2cm}|p{2cm}|}
        \hline
        \textbf{Data set} & \textbf{Training Time \newline per Epoch (s)} & \textbf{Peak GPU  \newline Memory (GB)}\\ \hline
        w/CC & 1922 & 17.86 \\ \hline
        w/o CC & 7416 & 47.54\\ \hline
    \end{tabular}
\end{table}

\begin{table}[ht]
    \centering
     \caption{Comparison of \textbf{Inference time(s)} / \textbf{GPU memory usage(GB)} between NC-PDNet with and without coil compression (w/CC and w/o CC, resp.) for 12-ch. and 32-ch. test volumes.   \label{tab:inference}}
     \vspace{0.cm}
    \begin{tabular}{|c|p{2.5cm}|p{2.5cm}|}
        \hline
        \textbf{} & \textbf{12-ch. volume} & \textbf{32-ch. volume}\\ \hline
        w/CC & 2.91(s) / 4.39(GB) & 4.95(s) / 5.49(GB) \\ \hline
        w/o CC & 7.75(s) / 7.57(GB) &21.1(s) / 19.74(GB)\\ \hline
    \end{tabular}   
\end{table}

Enabling coil compression reduces memory footprint and computational demand  as it is shown in Tab.~\ref{tab:training}, however training for more epochs is required to attain performance comparable to a model trained without coil compression. Importantly, coil compression does not compromise the structural integrity of the reconstructed images, although it may lead to a decrease in average PSNR. Additionally, using coil compression accelerates inference by approximately 2.66 times and 4.26 times for 12-channel and 32-channel volumes, respectively~(see Tab.~\ref{tab:inference}).

\section{DISCUSSION and CONCLUSION}
In this work, we extend NC-PDNet, a density-compensated unrolled neural network, to support the reconstruction of multi-coil 3D non-Cartesian k-space data. Our retrospective comparative study demonstrated that the unique design of the GoLF-SPARKLING sampling trajectories enables the recovery of higher-quality images compared to other sampling patterns. Presenting a benchmark of GoLF-SPARKLING's performance against commonly used non-Cartesian readouts in the literature is essential before advancing to further experiments. Our results indicated that while disabling coil compression provides enhanced performances in PSNR, this improvement has little impact on final image quality. We conclude that coil compression training in a channel-agnostic set-up is both practical and memory-efficient, achieving performance levels comparable to more complex training configurations. We reconstruct a 6x accelerated 1 mm isotropic 32-channel whole-brain 3D image in 4.95 seconds with a GPU memory usage of just 5.49 GB, demonstrating the clinical viability of this approach.
A key future direction is to validate NC-PDNet's performance on prospectively undersampled GoLF-SPARKLING data and assess its scalability in more challenging imaging setups, notably at higher resolution.

\section{Compliance with ethical standards}
\label{sec:ethics}

This study was conducted retrospectively using human subject data made available in open access \cite{beauferris2022multi}. Ethical approval was not required as confirmed by
the license attached with the open access data.

\section{Acknowledgments}
This work was granted access to the CCRT High-Performance Computing (HPC) facility under the Grant CCRT2024-tanaasma awarded by the Fundamental Research Division (DRF) of CEA. This work was granted access to IDRIS' HPC resources under the allocation 2023 AD011011153R3 made by GENCI. The concepts and information presented in this abstract are based on research results that are not commercially available. Future availability cannot be guaranteed.
\label{sec:acknowledgments}
\bibliographystyle{IEEEbib}
\bibliography{strings,refs}

\end{document}